\documentstyle[twocolumn,aps,epsfig]{revtex}
\begin{document}
 \wideabs{\title{Nonseparability and squeezing of continuous
polarization variables}

\author{Natalia Korolkova and Rodney Loudon}

\address{School of Physics and Astronomy, University of St. Andrews, North Haugh, St. Andrews KY16 9SS, UK}

\date{\today}
\maketitle \begin{abstract} {The impact of the operator-valued
commutator on nonclassical properties of continuous polarization
variables is discussed. The definition of polarization squeezing
is clarified to exclude those squeezed states which do not contain
any new physics beyond quadrature squeezing. We present a
consistent derivation of the general nonseparability criterion for
the continuous variables with the operator-valued commutator, and
apply it to the polarization variables.}\end{abstract} }

Within the framework of the quantum continuous variables,
nonclassical polarization states have recently attracted a
particular interest due to their compatibility  with the spin
variables of atomic systems and due to their simple detection
schemes \cite{polzik,korolkova,bowenSq,bowenEnt}. The relevant
continuous polarization variables are hermitian Stokes operators
(see \cite{korolkova} and references therein):
\begin{eqnarray}
\hat S_0 &=& \hat a_x^\dagger \hat a_x +  \hat a_y^\dagger \hat
a_y, \qquad \hat S_1 = \hat a_x^\dagger \hat a_x - \hat
a_y^\dagger \hat a_y , \nonumber\\ \hat S_2 &=& \hat a_x^\dagger
\hat a_y + \hat a_y^\dagger \hat a_x, \qquad  \hat S_3 = i \left(
\hat a_y^\dagger \hat a_x - \hat a_x^\dagger \hat a_y
\right)\label{S3}
\end{eqnarray}
where the $\hat a_x$ and $\hat a_y$ denote the bosonic photon
destruction operators associated with the $x$ and $y$ orthogonal
polarization modes. The Stokes operator $\hat S_0$ commutes with
all the others. The operators  $\hat S_j, j\neq 0$ obey the
commutation relations of the SU(2) Lie algebra:
\begin{eqnarray}
 [ \hat S_k, \hat S_l  ] = \epsilon_{klm} \ 2i\hat S_m,
\qquad k,l,m = 1,2,3. \label{commutator}
\end{eqnarray}
Simultaneous exact measurements of these Stokes operators are thus
impossible in general and their variances are restricted by the
uncertainty relations:
\begin{eqnarray}
  V_2V_3 \ge  \vert \langle \hat S_1 \rangle  \vert ^2, \
  V_3V_1 \ge  \vert \langle \hat S_2 \rangle  \vert ^2, \
  V_1V_2 \ge  \vert \langle \hat S_3 \rangle  \vert ^2 \label{Hei-S}
\end{eqnarray}
where $V_j=\langle \hat S_j^2\rangle - \langle \hat S_j\rangle^2$
is a shorthand notation for the variance of the quantum Stokes
parameter $\hat S_j$. The angle brackets denote expectation values
with respect to the state of interest.

Within the last few years, successful generation of polarization
squeezed \cite{polzik,bowenSq,heersink} and polarization entangled
\cite{bowenEnt,gloeckl} states has been reported. The respective
definitions of polarization squeezing \cite{polzik,chirkin} and
entanglement \cite{korolkova} were formulated. However, the subtleties arising due to the $q$-number,
i.e. operator-valued, commutator (cf. Eq.~(\ref{commutator})) have
not been satisfactorily discussed yet and a consistent derivation
of criteria for continuous variable polarization entanglement is
missing. To provide the answers to these open questions is the aim
of this paper.

A state is called {\it polarization squeezed} if:
\begin{eqnarray}
  V_k < |\langle \hat{S}_l \rangle | < V_m , \qquad k \neq l \neq m =1,2,3.
  \label{eqn:polsq_def}
\end{eqnarray}
The important difference between quadrature squeezing and polarization squeezing
is the discrepancy between coherent and minimum uncertainty states
for the latter. A coherent polarization state  is defined as a
quantum state with both polarization modes having a coherent
excitation $\alpha_{x}, \alpha_{y}$: $\psi_{coh} = \vert \alpha_x
\rangle_x \vert \alpha_y \rangle_y$. The quantum uncertainty of
such a state is equally distributed between the Stokes operators
and their variances are all equal to $V_j=V^{coh} = \vert \alpha_x
\vert^2+ \vert \alpha_y \vert^2=\langle \hat{n}
\rangle$. In analogy to quadrature squeezing, $V_j < V^{coh}$ 
seems at first glance to be a natural definition for polarization
squeezing. However, due to the SU(2) commutation algebra, a coherent
polarization state is not a minimum uncertainty state for all
three Stokes operators simultaneously. This was  known for atomic
states, i.e. for spin coherent states \cite{radcliffe}  and
angular momentum coherent states \cite{atkins}. The construction
of the minimum uncertainty product for the SU(2) algebra and the
properties of atomic coherent states were broadly studied around
early 70's \cite{radcliffe,atkins,jackiw,arecchi}. Although a
polarization state with a sub-shot-noise variance $V_j < V^{coh}$
is always a non-classical state, it implies nothing more than
conventional quadrature or single-mode squeezing observed through
the measurement of the Stokes parameter.

It is interesting to establish the relation 
of polarization squeezing to two-mode squeezing, i.e. quadrature entanglement. For two-mode squeezing, the
nonclassical correlations are created between two spatially separated modes. For polarization squeezing, quantum correlations are created between two orthogonal polarization modes. However, 
by the appropriate choice of variables and basis, the correlations within a two-mode system can be redistributed so that polarization squeezing is transformed into two-mode squeezing and vice versa. This effect was already observed in the experiments: Polarization squeezing and quadrature entanglement were observed in the same nonlinear system
of cold 4-level atoms, depending of the choice of the mode basis
\cite{giacobino}. Furthermore, the two schemes to generate continuous variable (CV) polarization entanglement have proven to be equivalent:  superimposing two polarization squeezed beams on a beam splitter
\cite{korolkova,gloeckl} or  overlapping two quadrature
entangled beams with an orthogonally polarized coherent beam each
\cite{bowenEnt}.   As an example of a basis transformation which translates both two-mode effects into each other let us view quadrature entanglement in terms of new variables having the mathematical form of Stokes operators:
\begin{eqnarray}
\hat s_0 &=& \hat a_A^\dagger \hat a_A +  \hat a_B^\dagger \hat
a_B, \qquad \hat s_1 = \hat a_A^\dagger \hat a_A - \hat
a_B^\dagger \hat a_B , \nonumber\\ \hat s_2 &=& \hat a_A^\dagger
\hat a_B + \hat a_B^\dagger \hat a_A, \qquad  \hat s_3 = i \left(
\hat a_B^\dagger \hat a_A - \hat a_A^\dagger \hat a_B
\right)\label{newS3}
\end{eqnarray}
where $A$ and $B$ are the two output spatially separated beams, the quadrature entangled beams.
To be specific, suppose that the quadrature entanglement emerges in the
interference of two amplitude-squeezed beams with equal squeezing $V^+$
and coherent amplitude $\alpha$ on a beam splitter \cite{silberhorn}, where the input amplitude squeezing is quantified by the variances $V^+<1<V^-$ of the quadrature operators $\hat X^+_{A,B} = \hat a_{A,B}^\dagger + \hat a_{A,B}$
and $\hat X^-_{A,B} = i ( \hat a_{A,B}^\dagger - \hat a_{A,B})$.  The variances
of the new "Stokes" operators $\hat s_j$ of Eq.~(\ref{newS3}) for the noncommuting pair $\hat s_1, \hat s_3$ are equal
to
\begin{eqnarray} {\rm v}_1 = 2\alpha^2V^- > \vert\langle  \hat s_2
\rangle\vert, \qquad
 {\rm v}_3 = 2\alpha^2V^+ < \vert\langle  \hat s_2
\rangle\vert .
\end{eqnarray}
Thus quadrature entanglement with anti-correlated amplitudes and
correlated phases exhibits  squeezing in $\hat s_3$.

Along with polarization squeezing, CV polarization entanglement \cite{korolkova} has proven to be a useful tool in quantum communication. There
is no unique criterion to quantify CV entanglement in general, in
particular for mixed states. For the generalization and comparison of different sum and product entanglement criteria for CVs with a $c$-number commutator see, e.g., Ref. \cite{giovannetti}. The formulation of 
the EPR criterion 
\cite{reid} and of the
nonseparability criterion 
\cite{duan} for the Stokes operators was presented in Ref.~\cite{korolkova} and further elaborated in Ref.~\cite{bowenEnt} on the basis of the generalized Heisenberg
relation. However, 
a consistent derivation of the nonseparability condition was not given. In the following, the general expression for the nonseparability criterion using CVs with an
{\it operator-valued} commutator is derived and the use of the generalized Heisenberg uncertainty relation is justified and emphasized. 

The standard derivation
\cite{jackiw,merzbacher} of the uncertainty relation for operators
$\hat A$ and $\hat B$ uses the Schwarz inequality in the form
\begin{eqnarray}
V_A V_B \ge&  \vert  \langle \Delta \hat A \Delta \hat
B
 \rangle  \vert ^2, \label{schwarz} 
 \end{eqnarray}
where $\Delta \hat A = \hat A - \langle \hat A \rangle$, $\Delta
\hat B = \hat B - \langle \hat B \rangle$ 
and the quantities on the left of (\ref{schwarz}) are the
variances  of the relevant operators (cf. Eq. \ref{Hei-S}).
The basic uncertainty relation can be written in a variety of
forms, for example \cite{jackiw}:
\begin{eqnarray}
V_AV_B \ge  \frac{1}{4} \left \vert \left \langle \left \lbrace
\Delta \hat A, \Delta \hat B \right \rbrace \right \rangle \right
\vert ^2 + \frac{1}{4} \left \vert \left \langle \left \lbrack
\Delta \hat A, \Delta \hat B \right \rbrack \right \rangle \right
\vert ^2, \label{varAvarB-com-anti-com}
\end{eqnarray}
where the anticommutator and commutator of the two operators are
defined by
\begin{eqnarray}
 \left \lbrace
\Delta \hat A, \Delta \hat B \right \rbrace & = & \hat A \hat B +
\hat B \hat A -
2 \langle \hat A \rangle \langle \hat B \rangle , \nonumber \\
\left \lbrack \Delta \hat A, \Delta \hat B \right \rbrack & = &
\hat A \hat B - \hat B \hat A = \lbrack \hat A, \hat B\rbrack.
\label{com-anti-com}
\end{eqnarray}

For {\it noncommuting operators}, both contributions on the right
of (\ref{varAvarB-com-anti-com}) are in general positive nonzero
quantities and the strongest inequality is obtained when both are
retained. However, valid but weaker inequalities result when one
or other of the contributions is neglected. Thus, removal of the
anticommutator term leads to the frequently-used form
\cite{merzbacher}
\begin{eqnarray}
V_AV_B \ge  \frac{1}{4} \left \vert \left \langle \left \lbrack
\Delta \hat A, \Delta \hat B \right \rbrack \right \rangle \right
\vert ^2. \label{un-rel-normal}
\end{eqnarray}
This provides the strongest available inequality when $\langle
 \lbrace \Delta \hat A, \Delta \hat B \rbrace
\rangle$ vanishes. The commutator $\lbrack \Delta \hat A, \Delta
\hat B  \rbrack$  is often a $c$-number as, for example, with the
position and momentum operators. In this case the variance
product has a universal state-independent lower limit,
 \begin{eqnarray}
V_AV_B \ge  \frac{1}{4} \left \vert  \left \lbrack \Delta \hat A,
\Delta \hat B \right \rbrack  \right \vert ^2.
\label{un-rel-c-number}
\end{eqnarray}
However, the commutator is sometimes an operator quantity, or
$q$-number, and the anticommutator $\lbrace \Delta \hat A, \Delta
\hat B \rbrace$  is usually a $q$-number. Both contributions on
the right of (\ref{varAvarB-com-anti-com}) then depend on the
state of the system
 and there is no reason to remove any of them. 
The inequality in (\ref{un-rel-normal})
remains valid but the full form in (\ref{varAvarB-com-anti-com})
provides a stronger inequality with a higher minimum value of the
variance product. There is no universal uncertainty relation in
such cases, as in the examples of the angular momentum operators
and of the Stokes-parameter operators considered here.

 The derivation of the nonseparability criterion for CV position
$x$ and momentum $p$ having a $c$-number commutator \cite{duan} considers an overall system composed
of two subsystems, $c$ and $d$, described by operators
\begin{eqnarray}
 \hat A = \vert a \vert \hat x_c +
\frac{1}{a}\  \hat x_d, \qquad \hat B = \vert a \vert \hat p_c -
\frac{1}{a} \ \hat p_d ,\label{comb-variabl}\\
\lbrack\hat x_i , \hat p_j  \rbrack = i \delta_{ij} \ (i,j=c,d),
\quad \lbrack\hat A , \hat B  \rbrack = i \left ( a^2 -
\frac{1}{a^2}\right ). \label{comb-variabl-comm}
\end{eqnarray}
The restrictions on the sum of the two
variances are direct consequences of the uncertainty relation: With the use of (\ref{schwarz}) and the Cauchy inequality $V_A^2 + V_B^2 \ge 2 V_A V_B $  it
follows that
\begin{eqnarray}
V_A + V_B \ge 2  \vert  \langle \Delta \hat A \Delta
\hat B  \rangle  \vert.  \label{var-sum-scwarz}
\end{eqnarray}
Thus, with the Heisenberg uncertainty relation taken in the form
(\ref{un-rel-c-number}), {\it all} states must satisfy
\begin{eqnarray}
V_AV_B \ge \frac{1}{4} \left ( a^2 - \frac{1}{a^2}\right )^2 \
{\rm and } \  V_A + V_B \ge \left \vert a^2 - \frac{1}{a^2}\right
\vert . \label{var-prod-sum}
\end{eqnarray}
It is shown in \cite{duan} that {\it separable} states of the two
subsystems must satisfy the stronger inequality
\begin{eqnarray}
V_A + V_B \ge  a^2 + \frac{1}{a^2} .
\end{eqnarray}
{\it Nonseparable} or {\it entangled} states thus exist in the
region defined by
\begin{eqnarray}
\left \vert a^2 - \frac{1}{a^2}\right \vert  \le V_A + V_B <  a^2
+ \frac{1}{a^2} , \label{nonsep-duan}
\end{eqnarray}
where the lower limit on the left comes from the development of
the Heisenberg uncertainty relation in (\ref{var-prod-sum}) and
the upper limit on the right comes from the nonseparability
criterion in \cite{duan} in its sufficient form.

We now rework the derivation of \cite{duan} for the
basic operator commutation relations more general than those given
in (\ref{comb-variabl}), (\ref{comb-variabl-comm}):
\begin{eqnarray}
 \hat A = \hat A_c +
\hat A_d, \qquad \hat B =  \hat B_c - \hat B_d,
\label{comb-variabl-gen}\\
\lbrack \hat A_c, \hat B_d \rbrack = \lbrack \hat B_c, \hat A_d
\rbrack =
0, \nonumber \\
\lbrack \hat A, \hat B \rbrack =  \lbrack \hat A_c, \hat B_c
\rbrack -  \lbrack \hat A_d, \hat B_d \rbrack.
\label{comb-var-comm-gen}
\end{eqnarray}
Here the nonzero commutators may themselves be opera-tors. The
uncertainty relations (\ref{var-prod-sum}) are generalized to 
 \begin{eqnarray}
 V_AV_B \ge \left \vert \left \langle \Delta \hat A_c \Delta \hat
 B_c\right \rangle - \left \langle \Delta \hat A_d \Delta \hat
 B_d\right \rangle \right \vert^2 ,  \label{var-sum-q-number1} \\  V_A + V_B \ge
 2 \left \vert \left \langle \Delta \hat A_c \Delta \hat
 B_c\right \rangle - \left \langle \Delta \hat A_d \Delta \hat
 B_d\right \rangle \right \vert . \label{var-sum-q-number}
 \end{eqnarray}
 Note that these relations reduce to those in (\ref{schwarz}) and
(\ref{var-sum-scwarz}) when there is only a single system,  $c$ or
$d$. By substitution of (\ref{comb-variabl-gen}) into (\ref{comb-variabl},\ref{var-sum-scwarz}), the Eq. (3,4) in \cite{duan} can be reworked for the pair of variables with the $q$-number commutator giving the {\it sufficient} nonseparability criterion. The main difference to the derivation of \cite{duan} is the replacement of the universal limit in (\ref{nonsep-duan}) by the state-dependent contribution containing the mean value of the operator-valued commutator (\ref{comb-var-comm-gen}) and the retainment of the state-dependent anticommutator contribution. {\it Nonseparable} or {\it entangled} states must then satisfy the condition
 \begin{eqnarray}
 2 \left \vert \left \langle \Delta \hat A_c \Delta \hat
 B_c\right \rangle - \left \langle \Delta \hat A_d \Delta \hat
 B_d\right \rangle \right \vert \le V_A + V_B < \nonumber \\
 2 \left \vert \left \langle \Delta \hat A_c \Delta \hat
 B_c\right \rangle \right \vert + 2 \left \vert \left \langle \Delta \hat
A_d \Delta \hat
 B_d\right \rangle \right \vert     ,  \label{nonsep-q-number}
      \end{eqnarray}
where the lower limit on the left comes from the development of
the Heisenberg uncertainty relation in (\ref{var-sum-q-number})
and the upper limit on the right comes from the generalization of
the {\it sufficient} nonseparability criterion  (see Appendix). A derivation of the nonseparability criterion  in its {\it necessary and sufficient} in the case of the $q$-number commutator still remains a challenge. The sufficient general {\it product} criterion was derived in Ref.~\cite{giovannetti}, where the standard form of the Heisenberg uncertainty relation was used to derive an upper limit for the product of two variances.

The expectation values that occur
in the limits of Eq.~(\ref{nonsep-q-number}) can be calculated either from the forms shown in
(\ref{nonsep-q-number}) or from the square root of the form shown
in (\ref{varAvarB-com-anti-com}). With this latter form, the
examples given below emphasize that the contributions of both the
anticommutator and the commutator must be retained to obtain the
most reliable and accurate limits. 

{\it Example 1: Korolkova et al.} The experiment proposed in
\cite{korolkova} uses bright light beams labelled $a$ and $b$,
each with $x$ and $y$ polarization components, all of which have
identical coherent amplitudes denoted by $\alpha$. Their
polarization squeezing properties are conveniently expressed in
terms of the variances of the quadrature operators defined by
\begin{eqnarray}
\hat X^+_{ax} =\hat a^\dagger _x +\hat a_x, \qquad \hat X^-_{ax}
=i \left ( \hat a^\dagger _x -\hat a_x \right )   \label{quadr}
\end{eqnarray}
with $\lbrack \hat X^+_{ax}, \hat X^-_{ax}\rbrack = 2i$ and
variances denoted $V^+_{ax}$, $V^-_{ax}$ (similarly for $ay$,
$bx$, and $by$). Consider the example with
\begin{eqnarray}
\hat A = \hat S_{1c} + \hat S_{1d}, \qquad \hat B = \hat S_{3c} -
\hat S_{3d},
\end{eqnarray}
where the subsystems  $c$ and $d$  refer to light beams produced
by interference of beams $a$  and  $b$, as described in
\cite{korolkova}. These have mean values of the Stokes parameters
given by
 \begin{eqnarray}
\langle \hat S_{1c} \rangle = \langle \hat S_{1d} \rangle =
\langle \hat S_{3c} \rangle = \langle \hat S_{3d} \rangle = 0,\nonumber\\
\quad \langle \hat S_{2c} \rangle = \langle \hat S_{2d} \rangle =
2\alpha^2. \label{means}
\end{eqnarray}

The various required quantities defined above are readily
calculated from expressions given in \cite{korolkova}. Thus,
\begin{eqnarray}
V_A=V_B=\alpha ^2 \left ( V^+_{ax} + V^+_{ay} + V^+_{bx} +
V^+_{by} \right ) {\rm and} \\
\left \langle \Delta \hat S_{1c}\Delta \hat S_{3c}\right \rangle =
\frac{1}{4}\alpha^2\left( V^+_{ax} - V^-_{ax}+ V^+_{ay} -
V^-_{ay}- \right .\nonumber \\ \left . V^+_{bx}+V^-_{bx}
-V^+_{by}+V^-_{by}\right) - 2i\alpha^2 = - \left \langle \Delta
\hat S_{1d}\Delta \hat S_{3d}\right \rangle
\end{eqnarray}
so that
\begin{eqnarray}
\left \vert \left \langle \Delta \hat S_{1c}\Delta \hat
S_{3c}\right \rangle \right \vert = \left \vert \left \langle
\Delta \hat S_{1d}\Delta \hat S_{3d}\right \rangle\right \vert =
\nonumber \\ \left \lbrace \frac{1}{16}\alpha^4\left( V^+_{ax} -
 V^-_{ax}+V^+_{ay} -
V^-_{ay}- V^+_{bx}+\right . \right .\nonumber \\ \left . \left .
V^-_{bx} -V^+_{by}+V^-_{by}\right)^2 + 4\alpha^4 \right \rbrace
^{1/2}. \label{corr-term}
\end{eqnarray}
This form shows the provenance of the two contributions to the
variance product (\ref{varAvarB-com-anti-com}), with the first
term in the square root coming from the anticommutator and the
second from the commutator, equal to $2 i\langle \hat S_2 \rangle$
in this example. In the simple case where the four modes making up
the polarization-squeezed beams all have equal quadrature
squeezing  $V^+$, $
V_A = V_B = 4\alpha^2 V^+$,
only the contribution of the commutator survives in
(\ref{corr-term}) to give
\begin{eqnarray}
\left \vert \left \langle \Delta \hat S_{1c}\Delta \hat
S_{3c}\right \rangle \right \vert = \left \vert \left \langle
\Delta \hat S_{1d}\Delta \hat S_{3d}\right \rangle\right \vert =
2\alpha^2 .
\end{eqnarray}
The range (\ref{nonsep-q-number}) for nonseparability therefore
becomes
\begin{eqnarray}
0 \le V^+ < 1. \label{end}
\end{eqnarray}
Note that the proper general form of the upper limit in the
entanglement criterion in (\ref{nonsep-q-number}) was not given in
\cite{korolkova} but the correct form of the specific result
(\ref{end}) was nevertheless derived there.

{\it Example 2: Bowen et al.} The experiment performed in
\cite{bowenEnt} also uses bright light beam subsystems labelled
$c$ and $d$ ($x$ and $y$ in notations of \cite{bowenEnt}), each of
which has $H$ and $V$ polarization components, with coherent
amplitudes $\alpha_H$ for both the $H$ components and $\alpha_V$
for both the $V$ components. Both $H$ components have quadrature
variances $V^+_H$ and  $V^-_H$, with a similar notation for the
common $V$ variances. The $H$ light beams are produced by
interference of primary quadrature-squeezed beams, then combined
with much more intense $V$ beams to form the $c$ and $d$
subsystems, in the manner described in \cite{bowenEnt}. The Stokes
parameter operators have the same general properties as outlined
above but with a generalization to allow for a phase difference
$\theta$ between the $H$ and $V$ components. For the performed
experiment with $\theta = \pi / 2$, the Stokes operators are
related to those used in \cite{korolkova}, which correspond to
$\theta = 0$, by
\begin{eqnarray}
\hat S_1(\frac{\pi}{2}) = \hat S_1 (0),  \ \  \hat
S_2(\frac{\pi}{2}) = - \hat S_3 (0), \ \  \hat S_3(\frac{\pi}{2})
= \hat S_2 (0). \label{conversion}
\end{eqnarray}
The expectation values of the  $\theta=\pi/2$ operators are
 \begin{eqnarray}
 \langle \hat S_1 \rangle = \alpha_H^2 - \alpha_V^2, \quad \langle
 \hat S_2 \rangle = 0 , \quad \langle \hat S_3 \rangle = 2
 \alpha_H \alpha_V,
 \label{meansBowen}\end{eqnarray}
in agreement with (\ref{means}) when $\alpha_H =\alpha_V=\alpha$
and the conversion (\ref{conversion}) is used.

The quantities needed for evaluation of the entanglement criteria,
again for the  $\theta = \pi / 2$ operators, are \cite{bowenEnt}
 \begin{eqnarray}
 V\left ( \hat S_{1c} \pm \hat S_{1d} \right ) = 2\alpha_V^2
 V^+_V, \nonumber \\  V\left ( \hat S_{2c} \pm \hat S_{2d} \right ) =
2\alpha_V^2
 V^-_H, \nonumber \\ V\left ( \hat S_{3c} \pm \hat S_{3d} \right ) =
2\alpha_V^2
 V^+_H ;\label{var-bowen}\\
 \left \vert \left \langle \Delta \hat S_1 \Delta \hat S_2 \right
 \rangle \right \vert = 2 \alpha_H \alpha _V, \nonumber \\
 \left \vert \left \langle \Delta \hat S_2 \Delta \hat S_3 \right
 \rangle \right \vert = \left \vert  \alpha_H^2 -  \alpha _V^2 \right
\vert, \nonumber \\
 \left \vert \left \langle \Delta \hat S_3 \Delta \hat S_1 \right
 \rangle \right \vert =  \alpha_H \alpha _V \left \vert V^+_H - V^+_V\right
\vert,
\label{corr-bowen}
\end{eqnarray}
with the same results for subsystems $c$  and  $d$. With use of
the form of variance product given in
(\ref{varAvarB-com-anti-com}), the first two expressions in
(\ref{corr-bowen}) result from the commutator, as is evident from
comparison with (\ref{commutator}) and (\ref{meansBowen}), and the
third from the anticommutator. 

The possibilities for entanglement with the three pairings of the Stokes
parameters are readily evaluated by substitution of the
expressions from (\ref{var-bowen}) and (\ref{corr-bowen}) in
(\ref{nonsep-q-number}). With the vertically-polarized input beams
assumed to be much brighter than the horizontally-polarized beams
$(\alpha_V \gg \alpha_H)$, the entanglement criterion is difficult
to satisfy for the pairs of Stokes operators  $\hat S_1, \hat S_2$
and $\hat S_3, \hat S_1$. However, the entanglement criterion for
the $\hat S_2, \hat S_3$ pair is the same as that for quadrature
entanglement and the corresponding polarization entanglement was
demonstrated experimentally \cite{bowenEnt}.

To conclude, the usual uncertainty relation in the form of
(\ref{un-rel-c-number}) is generalized to the equivalent forms in
(\ref{schwarz}) and (\ref{varAvarB-com-anti-com}) when the
commutator of the operators of interest is itself an operator, not
a $c$-number, and when the anticommutator of the operators is
nonzero \cite{jackiw,merzbacher}. The corresponding contributions
to the minimum variance product on the right of
(\ref{varAvarB-com-anti-com}) are then both state-dependent, in
contrast to the state-independent form of (\ref{un-rel-c-number}),
and there is no general reason to neglect either of them. The
generalization of the uncertainty relation also affects the range
of values for the variance sum in the usual nonseparability or
entanglement criterion \cite{duan} reproduced in
(\ref{nonsep-duan}), which is converted to the more general form
in (\ref{nonsep-q-number}). The retention of the anticommutator
contribution in (\ref{varAvarB-com-anti-com}) has the effect of
increasing the upper limit for entanglement on the right of
(\ref{nonsep-q-number}). The operator representations of the
Stokes polarization parameters provide examples of operator-valued
commutation relations, where the more general theory is needed for
the description of recently proposed \cite{korolkova} or performed
\cite{bowenEnt,gloeckl} experiments.

The authors thank G. Leuchs and T. C. Ralph for fruitful
discussions. The support of the Deutsche Forschungsgemeinschaft
and of the EU QIPC Project, No. IST-1999-13071 (QUICOV) is
gratefully acknowledged. R.L. is grateful to the Alexander von
Humboldt Foundation.

\appendix
\section{Derivation of the upper limit in Eq.~(22)}

Consider the variables $\hat A = \hat A_c + \hat A_d$ and $\hat B =  \hat B_c - \hat B_d$ (\ref{comb-variabl-gen}) of two subsystems $c$ and $d$. They obey the commutator relations  $\lbrack \hat A_c, \hat B_d \rbrack = \lbrack \hat B_c, \hat A_d
\rbrack =0$; $\lbrack \hat A_c, \hat B_c
\rbrack\neq 0$ and  $ \lbrack \hat A_d, \hat B_d
\rbrack \neq 0$ where the nonzero communtators  can be some nontrivial {\it operators}.
The variances of $\hat A$, $\hat B$  are defined as $V_Z =  \langle  (\Delta \hat Z  )^2  \rangle_\rho =
\langle \hat Z ^2 \rangle_\rho - \langle \hat Z \rangle^2_\rho$ with $Z=A, B $ and $\Delta \hat Z = \hat Z - \langle \hat Z \rangle$.\\

{\it Theorem.} For any separable state $\rho_{sep}$ the following inequality holds:
\begin{eqnarray}
 V_A + V_B \ge
 2 \left \vert \left \langle \Delta \hat A_c \Delta \hat
 B_c\right \rangle \right \vert + 2 \left \vert \left \langle \Delta \hat
A_d \Delta \hat
 B_d\right \rangle \right \vert      \label{sep-q-number}
      \end{eqnarray}  \\

{\it Proof.}
A separable quantum state $\rho_{sep}$ can be written as a convex decomposition
\begin{eqnarray}
\rho_{sep} = \sum_j\  p_j \rho_{jc}\otimes \rho_{jd}. \label{sep}
\end{eqnarray}
Using this decomposion we can directly compute the sum of the variances $V_A + V_B$. The averaging in the expressions below is performed over the product density matrix $\rho_{sep} = \sum_j\  p_j \rho_{jc}\otimes \rho_{jd}$. We obtain:
\begin{eqnarray}
V_A+V_B   =  \sum_j p_j\ \left ( \langle \hat A^2 \rangle_j + \langle \hat B^2 \rangle_j\right ) \nonumber \\  - \left ( \sum_j p_j \langle \rangle_j \right )^2 - \left ( \sum_j p_j \langle B\rangle_j \right )^2 \nonumber \\
 =  \sum_j p_j \left (  \langle  \hat A_c^2 \rangle_j + \langle\hat A_d^2 \rangle_j + \langle \hat B_c^2 \rangle_j
+ \langle \hat B_d^2 \rangle_j\right ) \nonumber \\
+ 2\left ( \sum_j p_j \langle A_c\rangle_j\langle A_d\rangle_j - \sum_j p_j \langle B_c\rangle_j\langle B_d\rangle_j \right )\nonumber \\  -  \left ( \sum_j p_j \langle A \rangle_j \right )^2 - \left ( \sum_j p_j \langle B\rangle_j \right )^2 = \nonumber 
\end{eqnarray}
\begin{eqnarray}
\sum_j p_j \left (  \langle (\Delta \hat A_c )^2 \rangle_j + \langle (\Delta \hat A_d )^2 \rangle_j + \langle (\Delta \hat B_c )^2 \rangle_j
+ \langle (\Delta \hat B_d )^2 \rangle_j\right ) \nonumber \\
\sum_j p_j \left (\langle\hat A\rangle^2_j + \langle\hat B \rangle^2_j \right ) - \left ( \sum_j p_j \langle A \rangle_j \right )^2 - \left ( \sum_j p_j \langle B\rangle_j \right )^2. 
\label{ineq}
\end{eqnarray}
Let us estimate the limits for the last two lines in (\ref{ineq}). We use the Schwarz inequality in the form (\ref{schwarz}) and $V_A^2 + V_B^2 \ge 2 V_A V_B$ and get:
\begin{eqnarray}
\sum_j p_j \left (  \langle (\Delta \hat A_c )^2 \rangle_j + \langle (\Delta \hat B_c )^2 \rangle_j + \langle (\Delta \hat A_d )^2 \rangle_j + \langle (\Delta \hat B_d )^2 \rangle_j
\right ) \nonumber \\
\ge 2 \left \vert \left \langle \Delta \hat A_c \Delta \hat
 B_c\right \rangle \right \vert + 2 \left \vert \left \langle \Delta \hat
A_d \Delta \hat
 B_d\right \rangle \right \vert.  \nonumber
\end{eqnarray}
Note that the application of the Schwarz inequality (\ref{schwarz}) corresponds to the use of the generalized uncertainty relation: Eq.~(\ref{schwarz})  is readily re-expressed in the form Eq.~(\ref{varAvarB-com-anti-com}) and the anti-commutator term is retained.
Furthermore, it can be easily shown \cite{duan} using the Cauchy-Schwarz inequality $(\sum_j p_j)(\sum_j p_j \langle \hat A \rangle^2_j) \ge (\sum_j p_j \vert \langle \hat A \rangle_j \vert )^2$ that the lower bound for the last line in (\ref{ineq}) is zero,
\begin{eqnarray}
\sum_j p_j \left (\langle\hat A\rangle^2_j + \langle\hat B \rangle^2_j \right ) \nonumber \\ - \left ( \sum_j p_j \langle A \rangle_j \right )^2 - \left ( \sum_j p_j \langle B\rangle_j \right )^2 \ge 0. 
\nonumber
\end{eqnarray}
Hence, for any separable state (\ref{sep}) the inequality (\ref{sep-q-number}) holds, which proves our statement.\\

It follows from Eq.~(\ref{sep-q-number}) and uncertainty relations (\ref{schwarz}), (\ref{varAvarB-com-anti-com}) that the nonseparable or entangled states have to satisfy Eq.~(\ref{nonsep-q-number}):
\begin{eqnarray}
 2 \left \vert \left \langle \Delta \hat A_c \Delta \hat
 B_c\right \rangle - \left \langle \Delta \hat A_d \Delta \hat
 B_d\right \rangle \right \vert \le V_A + V_B < \nonumber \\
 2 \left \vert \left \langle \Delta \hat A_c \Delta \hat
 B_c\right \rangle \right \vert + 2 \left \vert \left \langle \Delta \hat
A_d \Delta \hat
 B_d\right \rangle \right \vert     .  \label{nonsep-q-number-Ap}
      \end{eqnarray}
In contrast to Eqs.~(5-7) \cite{typo} in Ref.~\cite{giovannetti} the lower limit in (\ref{sep-q-number}) and hence the upper limit in Eqs.~(\ref{nonsep-q-number}), (\ref{nonsep-q-number-Ap}) does not depend on the particular form of the convex decomposition in (\ref{sep}). However, the lower bound for $V_A+V_B$ (\ref{sep-q-number}) and the limits in the nonseparability criterion Eq.~(\ref{nonsep-q-number}), (\ref{nonsep-q-number-Ap}) do depend on the quantum state under consideration. There is no universal separability limit for the sum or product of the two variances $V_A$, $V_B$ in the case of the operator-valued commutators $\lbrack \hat A_c, \hat B_c
\rbrack$ and  $ \lbrack \hat A_d, \hat B_d
\rbrack$. Nevertheless, the inequalities of Eqs.~(\ref{sep-q-number}), (\ref{nonsep-q-number}), (\ref{nonsep-q-number-Ap}) provide a sensible operational sufficient criterion for nonseparability which can be readily verified in an experiment (see examples in the paper).

\end{document}